Titel: Artificial Left Ventricle

Authors: Saeed Ranjbar [1,*], Ph.D, Tohid Emami Meybodi [2], MD, , Mahmood emami Meybodi [2], MD

**Research institute:**

1- Modarres Hospital, Institute of Cardiovascular Research, Shahid Beheshti University of Medical Science, Tehran, Iran.

2- Shahid sadoughi university of medical sciences, Yazd, Iran

**\*Corresponding Author:** Saeed Ranjbar

The full postal address of the corresponding author:

Modarres Hospital, Institute of Cardiovascular Research, Shahid Beheshti University of Medical Science, Tehran, Iran

E-mail: sranjbar@ipm.ir

Tel/FAX: +9821 22083106

# Artificial Left Ventricle

Saeed Ranjbar [1,*], Ph.D, Tohid Emami Meybodi [2], MD, , Mahmood emami Meybodi [2], MD

**Abstract**

This Artificial left ventricle is based on a simple conic assumption shape for left ventricle where its motion is made by attached compressed elastic tubes to its walls which are regarded to electrical points at each nodal .This compressed tubes are playing the role of myofibers in the myocardium of the left ventricle. These elastic tubes have helical shapes and are transacting on these helical bands dynamically. At this invention we give an algorithm of this artificial left ventricle construction that of course the effect of the blood flow in LV is observed with making beneficiary used of sensors to obtain this effecting, something like to lifegates problem. The main problem is to evaluate powers that are interacted between elastic body (left ventricle) and fluid (blood). The main goal of this invention is to show that artificial heart is not just a pump, but mechanical modeling of LV wall and its interaction with blood in it (blood movement modeling) can introduce an artificial heart closed to natural heart.

FIGURE 1

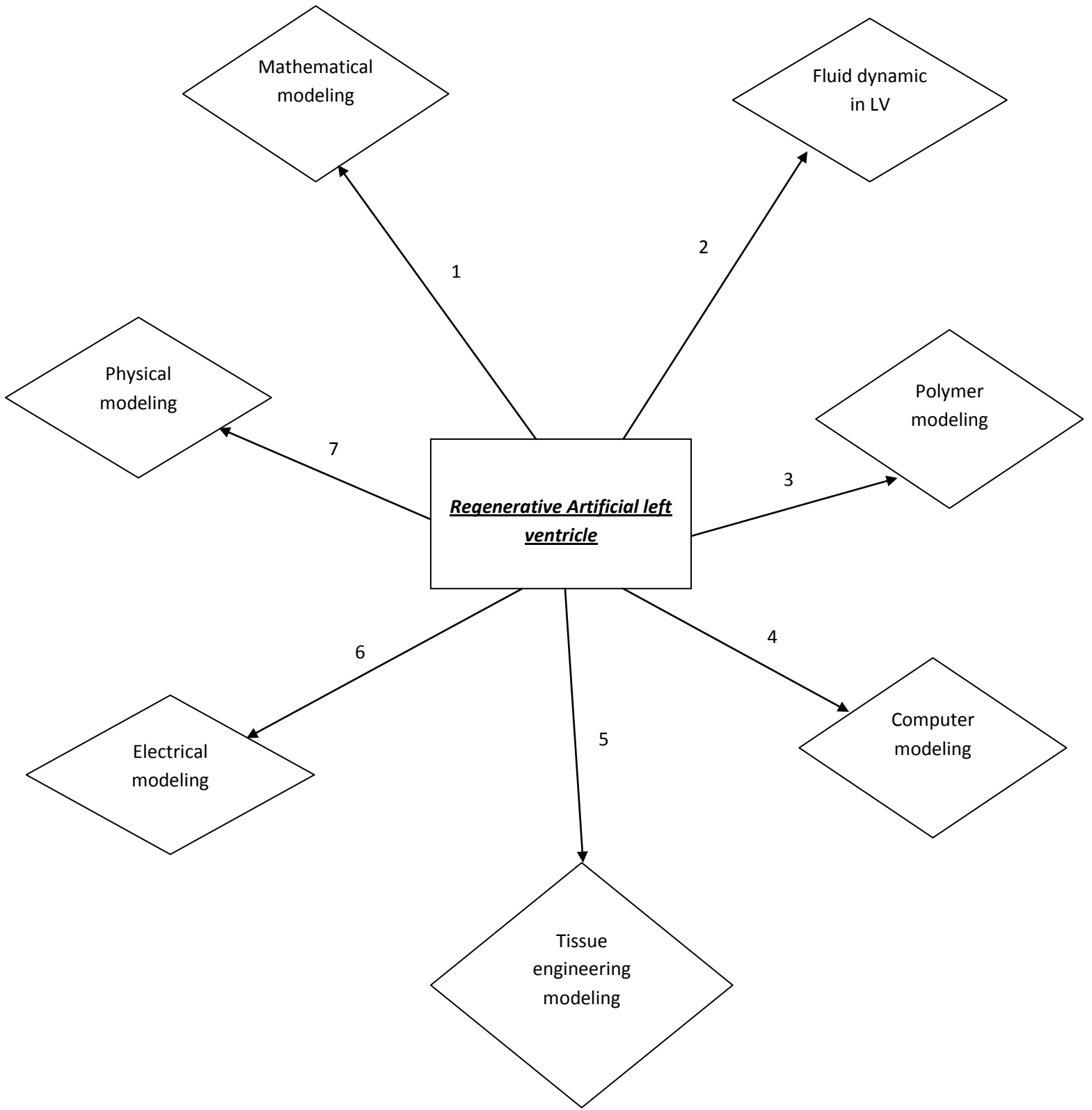

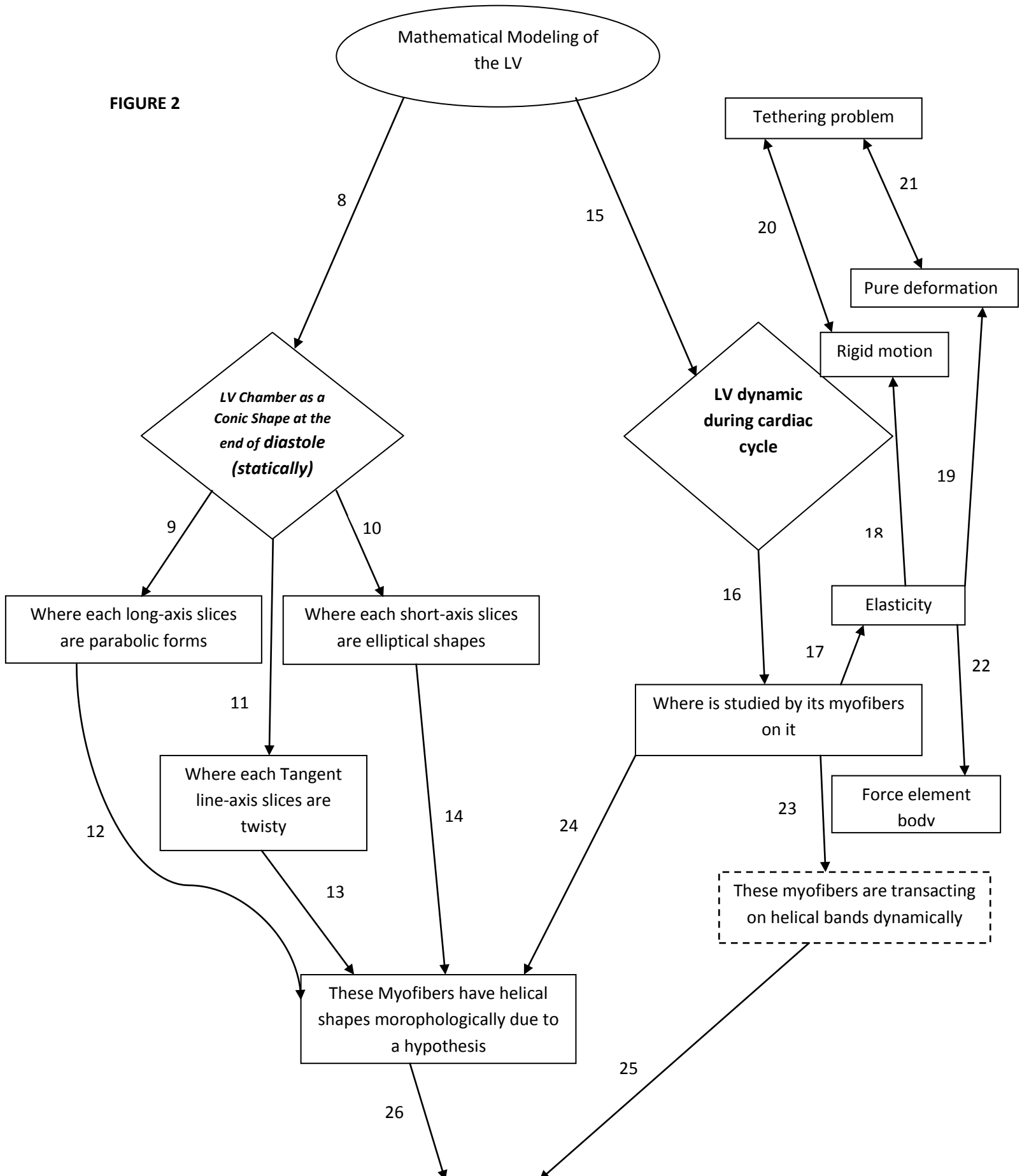

**FIGURE 2**

*These bands initiate from the posterior-basal region of the heart continues through the left ventricle free wall, reaches the septum, loops around the apex, ascends, and ends to the superior-anteriorsection of the left ventricle.*

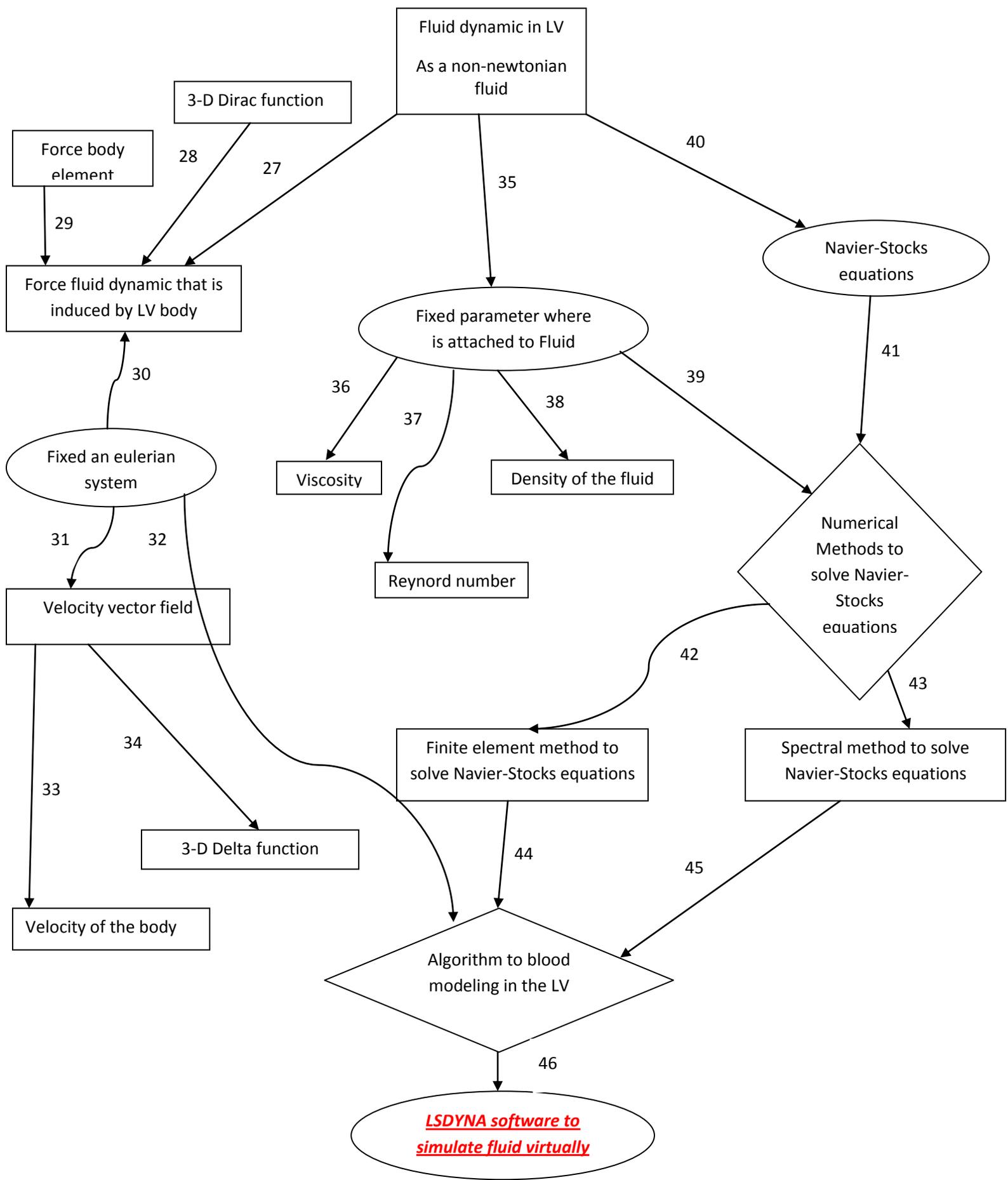

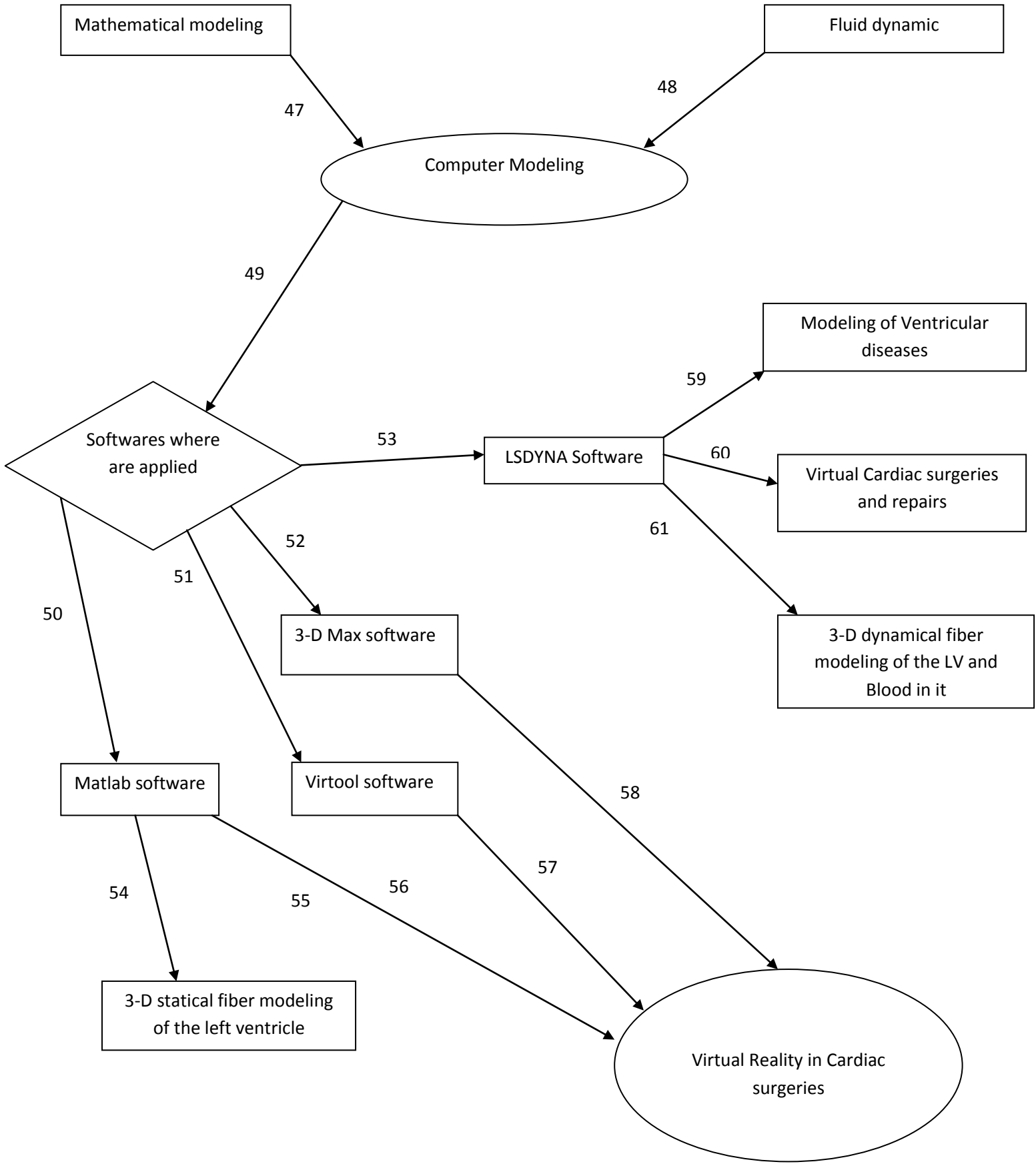

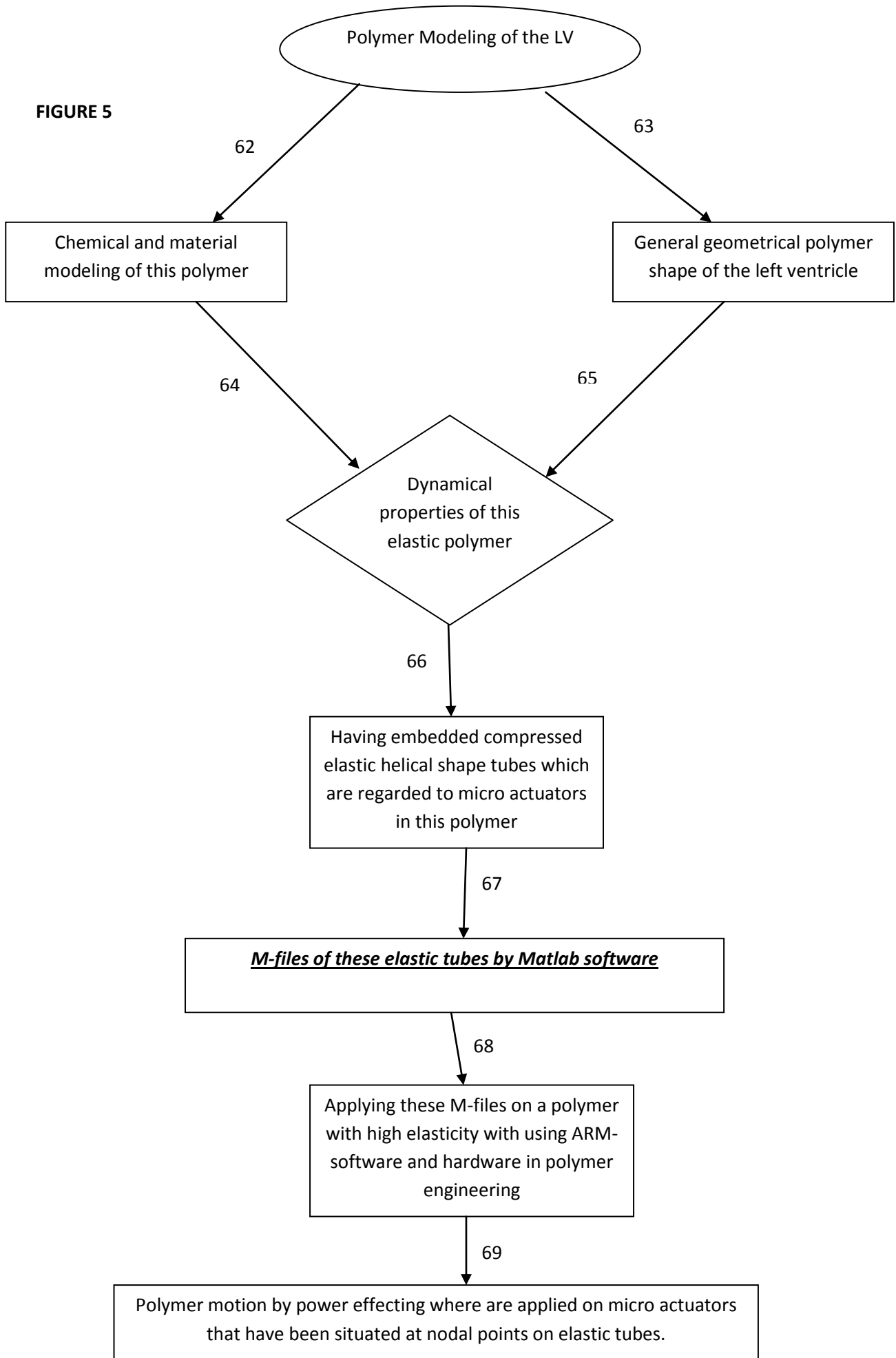

**FIGURE 5**

FIGURE 6

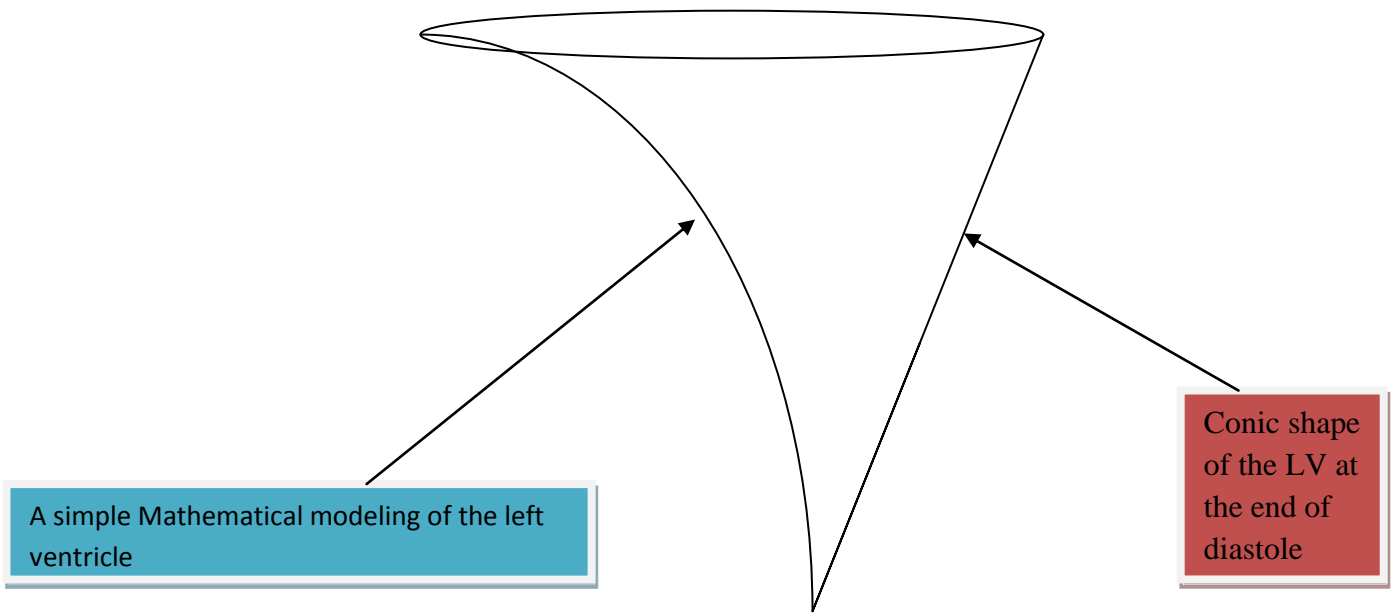

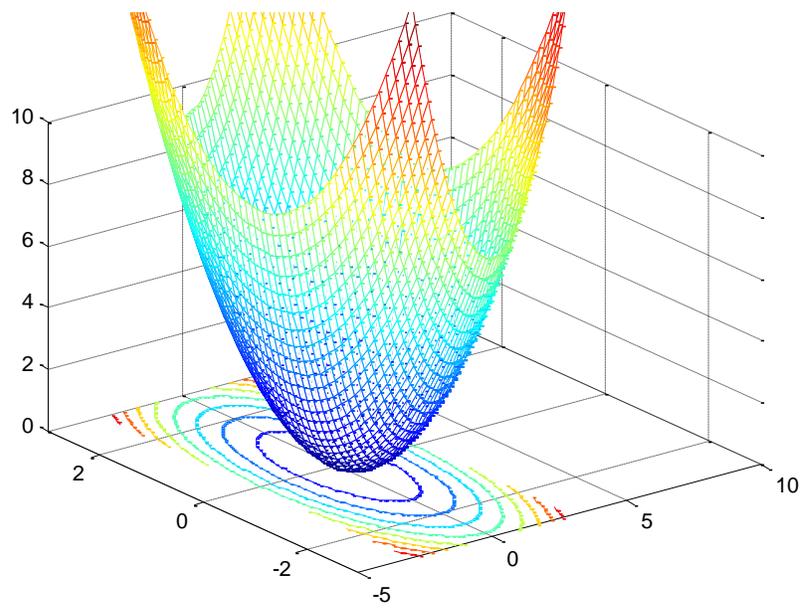

Figure 7

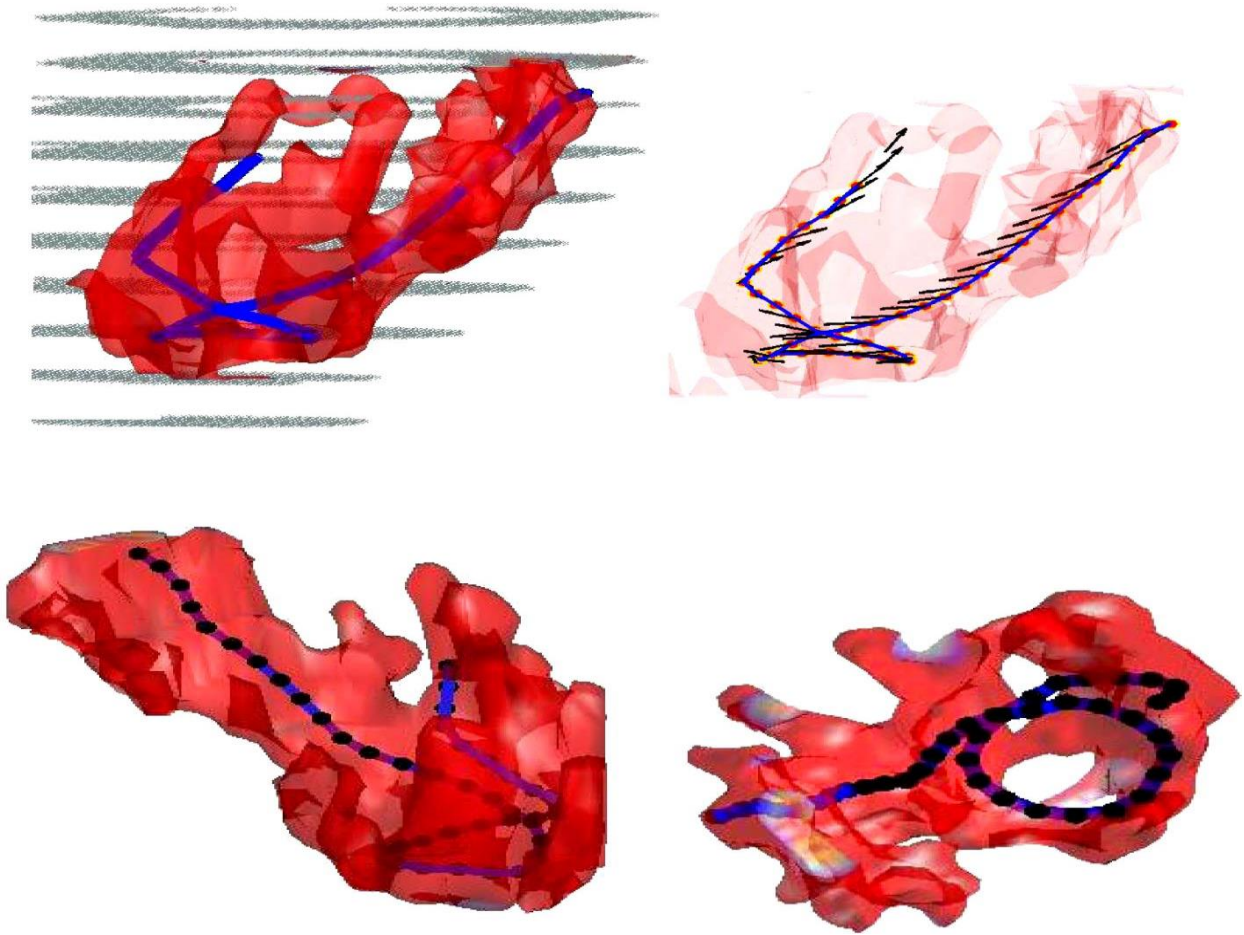

Figure 8

# FIGURE 9

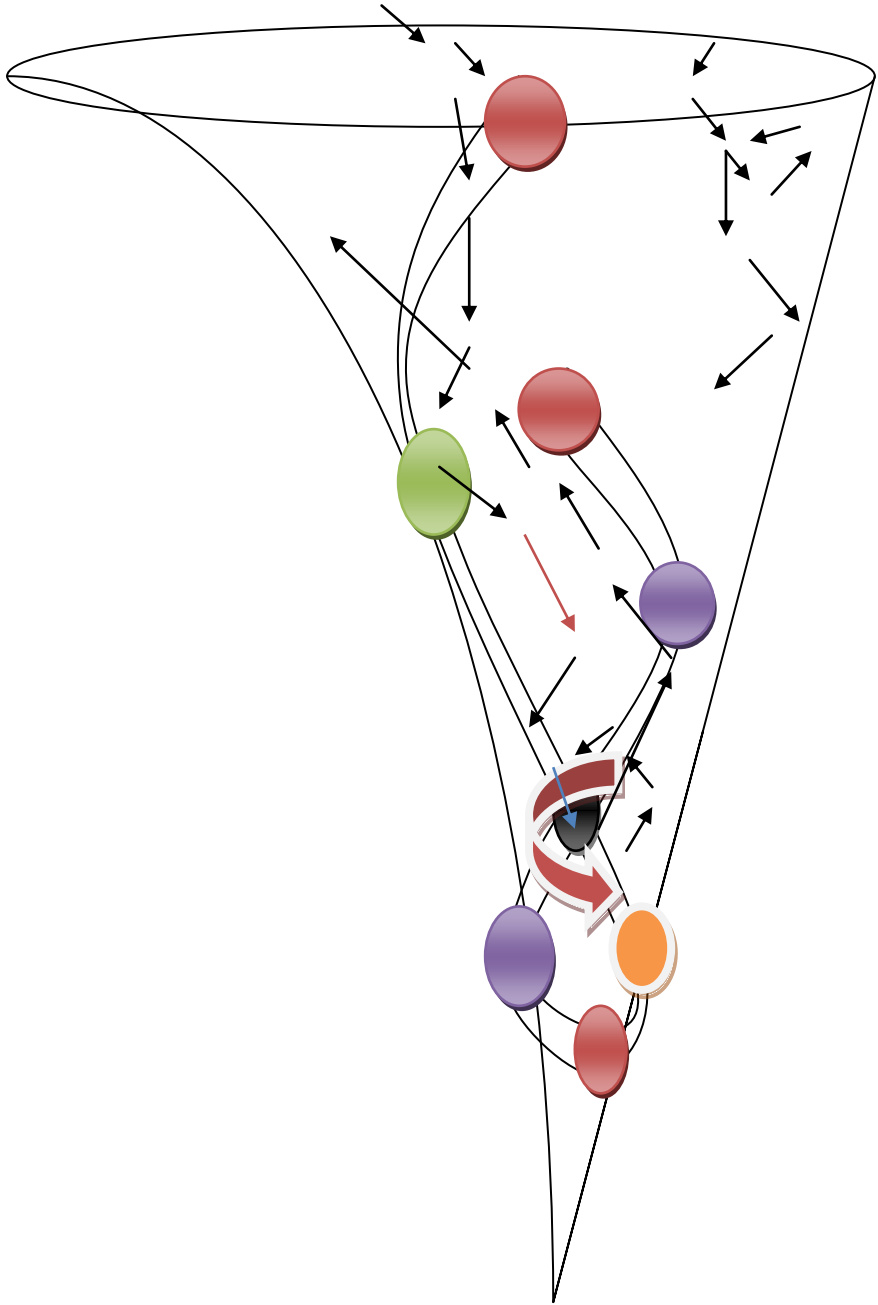

Having used sensors at each nodal point, blood force on the LV wall is considered.

FIGURE 10

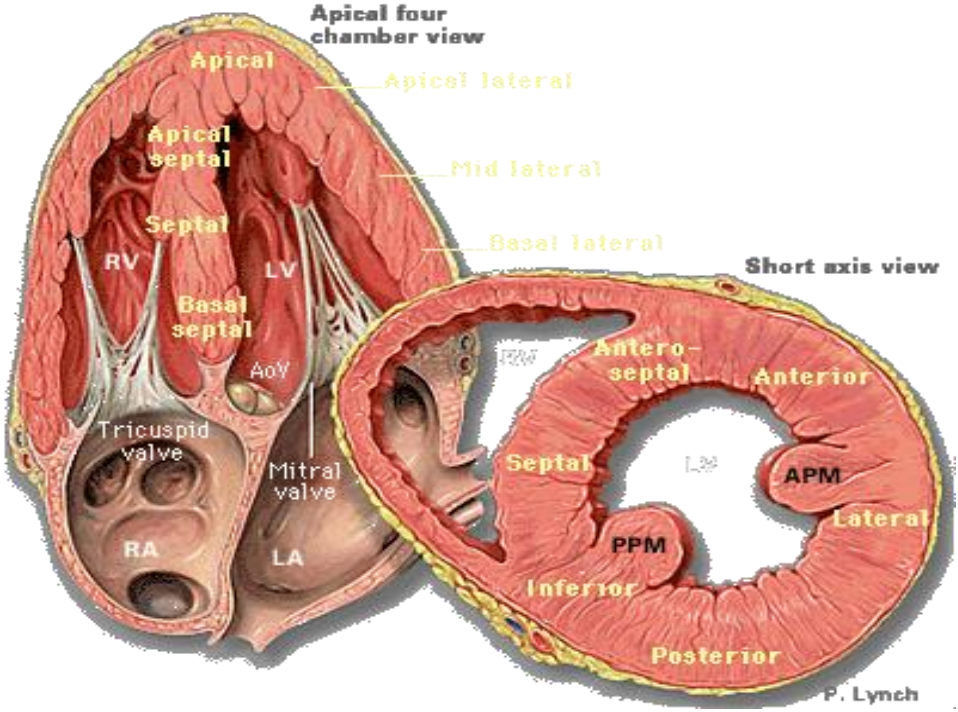

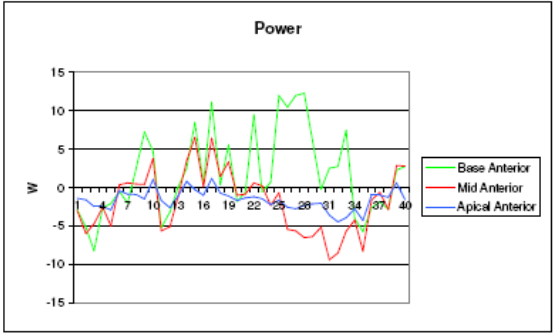
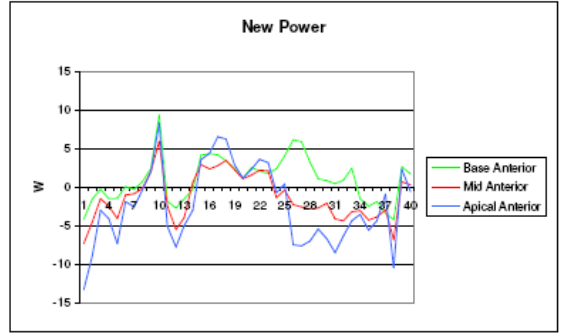
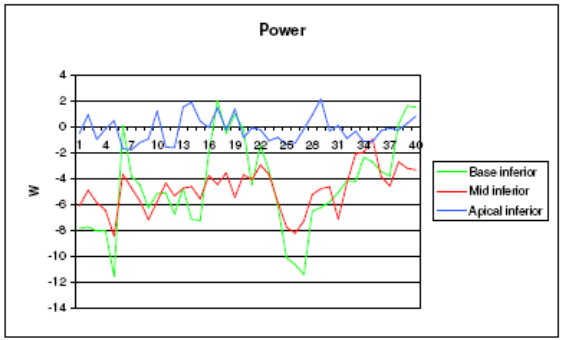
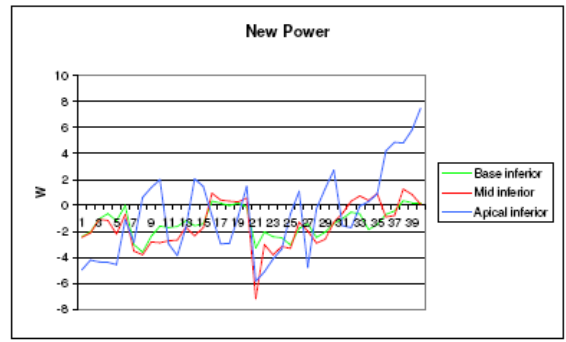
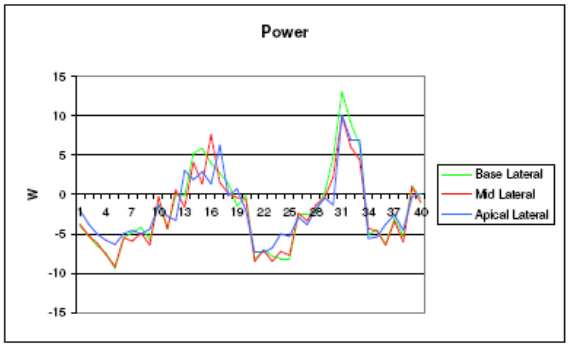
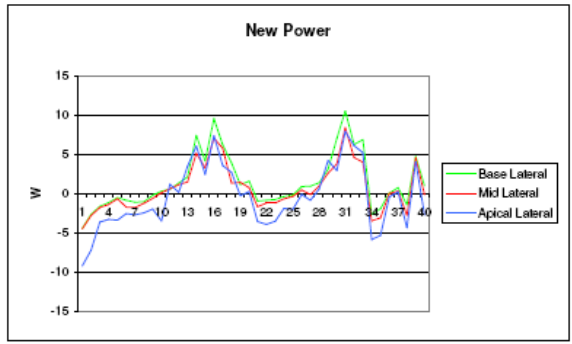
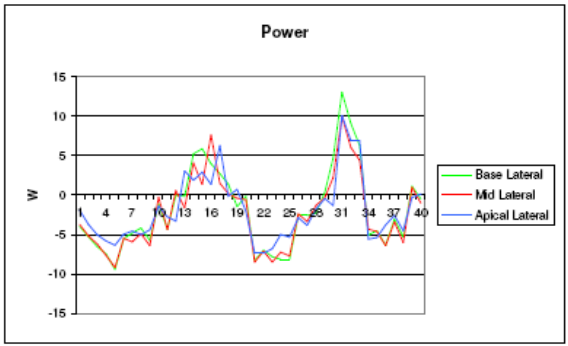
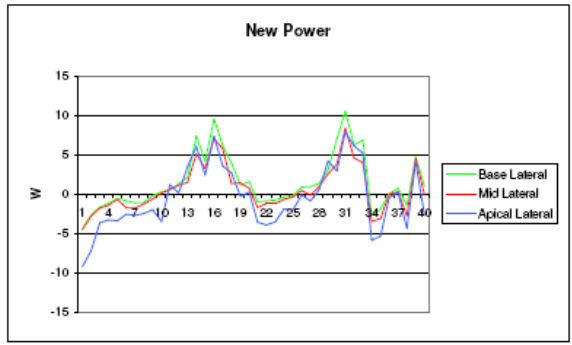

FIGURE 12

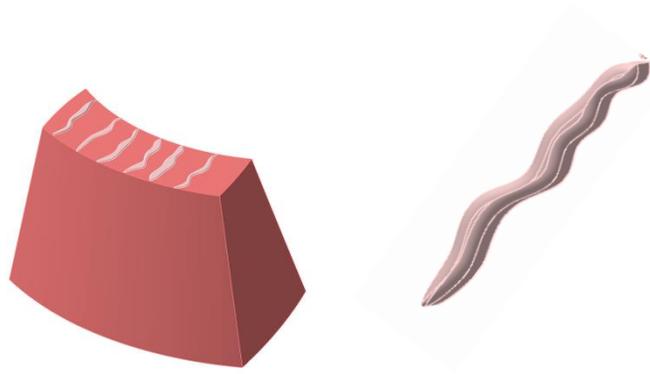

FIGURE 13

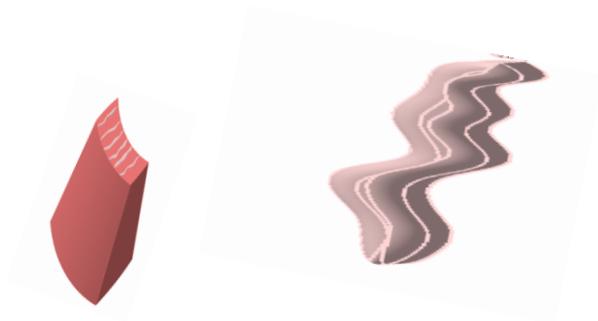

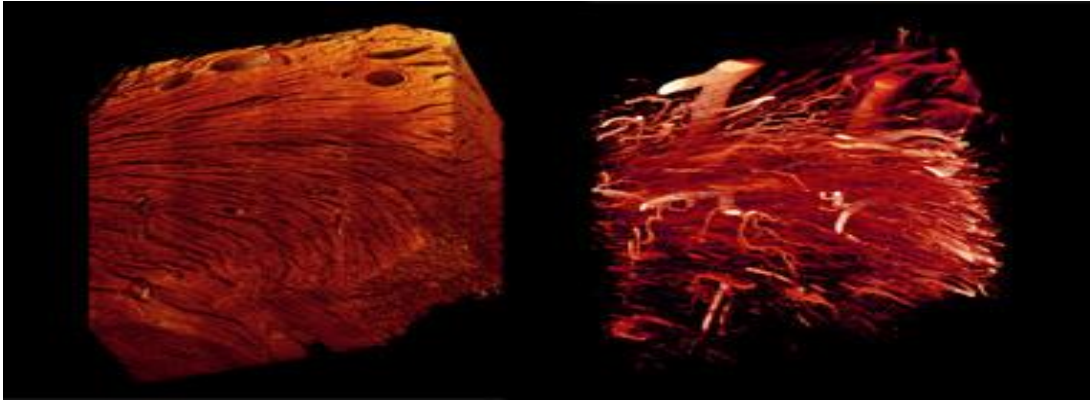
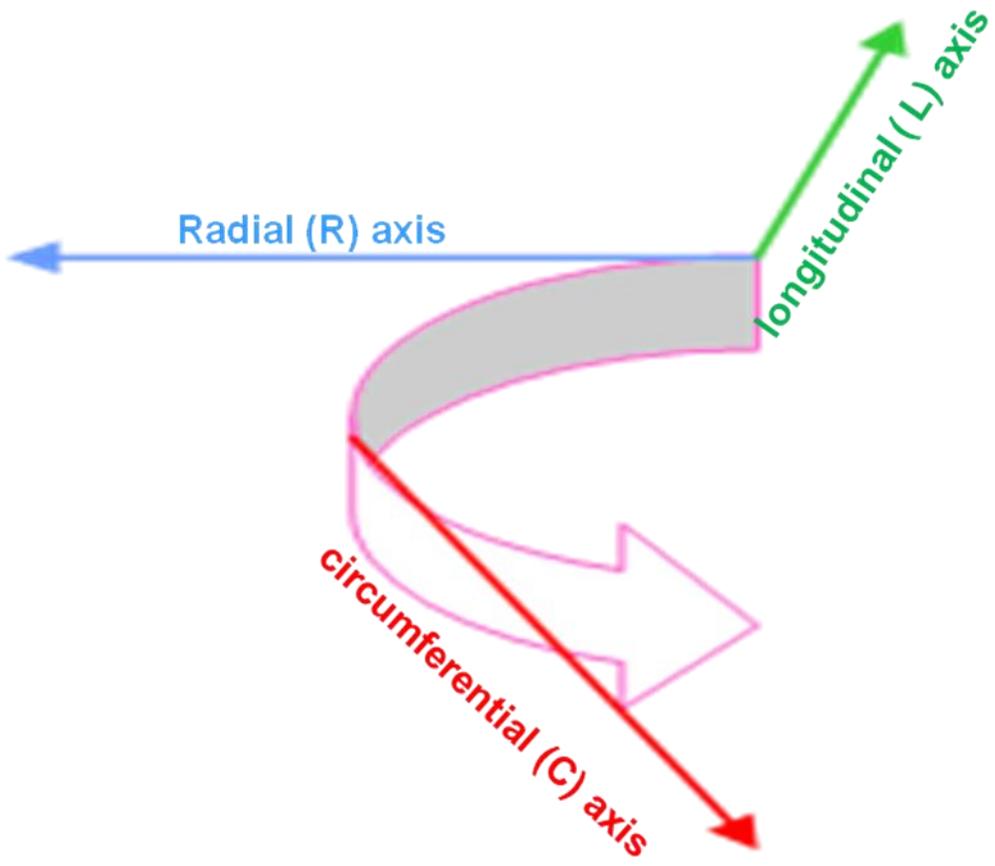

Figure 14

## *Background of the invention*

All recent artificial left ventricles are just miniature pumps. **The pump** Electric coils cause a rotor with an embedded magnet to spin. Fins on the rotor push Oxygen-rich blood (red arrows) from the heart through a tube in to the aorta. On October 27, 2008, French professor and leading heart transplant specialist Alain F. Carpentier announced that a fully implantable artificial heart will be ready for clinical trial by 2011 and for alternative transplant in 2013. It was developed and will be manufactured by him, Biomedical firm Carmat, and venture capital firm Truffle. The prototype uses electronic sensors and is made from chemically treated animal tissues, called "biomaterials", or a "pseudo-skin" of biosynthetic, microporous materials. Another US team with a prototype called 2005 MagScrew Total Artificial Heart, including Japan and South Korea researchers are racing to produce similar projects. But at this presentation the main goal is to advance recent attempts on artificial left ventricles by these point of views where using compressed elastic tubes with helical shapes toward microprous materials which are applied to nodal points of these elastic tubes. At this invention these elastic tubes are replaced with the fibers of the myocardium in the left ventricle where this is needed to find a good understanding of these fibers not only statically but also dynamically at the same time. A fiber in the myocardium means to follow the curve of the motion of a muscle volume sample in the myocardium from the end of diastole to the end of systole. Characterization of local and global contractile activities in the myocardium is essential for a better understanding of cardiac form and function. The spatial distribution of regions that contribute the most to cardiac function plays an important role in defining the pumping parameters of the myocardium like ejection fraction and dynamic aspects such as twisting and untwisting. In general Myocardial shortening and lengthening, tangent to the wall , and ventricular wall thickening are important parameters that characterize the regional contribution within the myocardium to the global function of the heart.

We calculated these parameters using vector bundles of myocardium deformation, which were captured through the displacement encoding with stimulated echoes in 50 normal volunteers. High spatial resolution of the acquired data revealed transmural changes of wall thickening and tangential shortening with high fidelity in beating hearts during a cycle. By filtering myocardium regions that showed a tangential shortening of < 0.23, we were able to identify the complete or a portion of a macrostructure composed of connected regions in the form of a helical bundle within the left ventricle mass. In this study, we present a represetative case that shows the complete morphology of a helical myocardial band due to a famous hypothesis about fibers transaction in the LV as well as cases that present ascending and descending portions of the helical myocardial band.

A better understanding the structure and function of the heart continues to challenge modern medicine and physiology. Despite the remarkable advances in the understanding of myocardial cell function and its complex molecularstructure, the role of cardiac sarcomeres and myofibers in generating contractile forces in the normal heart baffles us at all levels. A better understanding of the spatial distribution of regions with elevated contractile dynamics within the myocardium mass would perhaps help to resolve this issue. As it can also be observed in the morphology of the LV, our hypothesis is that if myofibers of the left ventricle form a gross macroscopic structure in the form of a band, its importance in the dynamics of the heart must be revealed in characterized regional and global function of the myocardium. Based on our hypothesis, such an anatomic structure should act like a physical pathway for the maximum transmission of systolic contractile force and should facilitate the spatial coordination of relaxation and diastolic recoil. The effective of this approach depends on the proper indentification of the cardiac regions that exhibit higher contractile activity and the nature of their linkage and inner connectivity. In this presentation, we implement an approach that enable us to map regions of intense and directional contractile activities within myocardium to examine the nature of gross morphology suggests that a band in the form of elleptical characterizes the most intense contractile activities within the ventricular mass.

The other assessment to gain a necessary understanding of the local contractility, in contrast with chemical and electrical reasons of the cardiac muscle contraction would be followed by a new notion of the intrinsic power of myocytes resulting in the sufficient energy to the cardiac motion and deformation. If we follow myofiber's transactions from the end diastole to the end systole, they would show us helical curves that intrinsic power is made to move myofibers on these bands. The other values must be calculated is the local power of the fluid (blood) which is transfered to the myocardium.

It would therefore be desirable to design a method and system to fully applying myofiber transactions in the myocardium and their interactions with the fluid through the LV for this assistant devise where it has the central role in the cardiac function.

## Summary

This invention includes a method and system that is applied to an elastic conic shape where here is a miniature of the left ventricle.

In fact our elastic material of the left ventricle is a plastic modeling of a simple mathematical modeling of the left ventricle in compared with its physical modeling.

The motion and deformation of the LV are forced by studying of the behaviors of fibers in LV, statically and dynamically. Our evidence not only in cardiac imaging but also in the mathematical modeling show that the most of these fibers move on helical bands. Here these fibers in the myocardium are replaced with compressed elastic tubes where are deformed to helical shapes. For each part of LV wall, they set somehow in the myocardium where their nodal points pass from the LV free wall, the mid of the septum, the apex, the mid of the lateral and the base of the anterior respectively.

Electrical sensors which are made by microprous materials are used at these nodal points to push elastic tubes on their helical curves. The powers which are recharged by these microprouses save from the evaluation of the power of the local myocardium at our LV mathematical modeling. And by these sensors and having used evaluated powers of the blood when comes in to the LV, we can observe effecting of the blood to body movement of the left ventricle.

## Brief Description of the Drawings

The drawings illustrate one preferred embodiment presently contemplated for carrying out the invention.

In the drawings:

FIG. **1** is a general flowchart of a regenerative artificial left ventricle where states what models are involved to pose a good structure of an artificial left ventricle near real left ventricle.

FIG. **2** is a flowchart of the mathematical modeling of the left ventricle function not only statically but dynamically at the same time where data are evaluated by electrocardiogram data in compare with the other cardiac imaging tools.

FIG. **3** is a flowchart of the fluid (the blood) dynamic as a non-newtonian fluid in the left ventricle where is studied as fluid-wall interaction. samples for solving partial differential equations come from Echo ( as one of tools in cardiac imaging by ultrasound method).

FIG. **4** and **5** are flowcharts of computer and polymer modeling at this invention.

FIG. **6** shows a simple mathematical conic assumption of the left ventricle at the end of diastole where is constructible by an elastic polymer.

FIG. **7** is a mesh file of this conic shape at the Matlab software.

FIG. **8** shows nodal points in a compressed elastic twisty tube. Micro actuators are applied at these points to transfer enough energy that we have explicitly computed these values of energy.

FIG. **9** is a scheme of using sensors at nodal points to evaluate the effect of the blood though the mitral valve and left ventricle.

FIG. **10** is a short-axis view of the left ventricle toward landmarks of experimental samples.

FIG. **11** is the diagrams of power local values of the myocardium where have been tested on 50 normal cases by Vivid 7 Echo.

FIG. **12** is a scheme of a muscle volume sample toward a myofiber on it in the myocardium at the end of diastole.

FIG. **13** is the same muscle volume sample after passing 0.2s from the end of diastole.

FIG. **14** is a scheme of a muscle volume sample in the myocardium with a distribution of its fibers in it. And a scheme of three vectors of deformations where is attached to this volume sample.

## ***Detailed Description of the prefferd of embodiment:***

Referring to FIG. **1**, the major components of a regenerative artificial left ventricle **1-6** incorporating the present invention are shown. The operation of this regenerative artificial left ventricle is locally designed mathematically to FIG. **2, 8** and **15** in compared with its real physical modeling of the left ventricle. Behavior of the left ventricle is basically divided to statical modeling of the LV **9-14** and to dynamical studying of the LV **16-25**. Having assumed LV as a conic shape **8** at the end of diastole and look at shapes of its sections **9-11**, we can gain a good understanding of the local geometrical shape of the myocardium at the left ventricle. These local mathematical shapes are attached to each muscle volume samples and the main theme is to follow these samples on the curves where they are the fibers of a deformation flat map **16-19**. In FIG. **14**, left ventricle can be realized as a vector bundle of deformation components $\varepsilon_{rr}$ radial strain $\varepsilon_{ll}$ longitudinal strain and $\varepsilon_{cc}$ Circumferential strain that are evaluated by Echo techniques for each landmarks at our samples. And the curves of myofibers transactions are iso-curves of a deformation map which is defined by the following way:

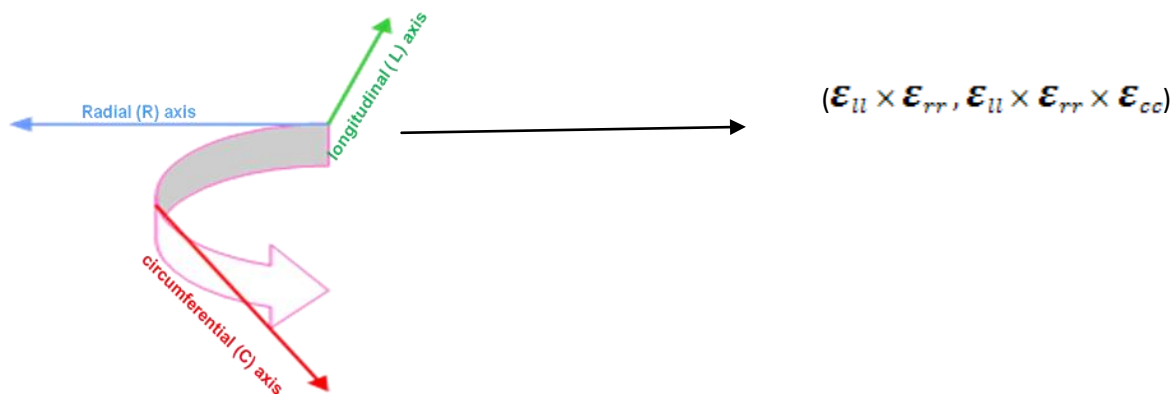

$$(\varepsilon_{ll} \times \varepsilon_{rr}, \varepsilon_{ll} \times \varepsilon_{rr} \times \varepsilon_{cc})$$

fibers (as an outputs of an algorithm where is running co-images of the deformations map) of this map have a twisty shape geometrically **12-14**, and fiber transactions on such these bands where are on iso-curves of the deformation map give us a fiber modeling of the myocardium at the left ventricle **23-25**. The advantage of this modeling is due to ultrasound method where is really accessible and our algorithm as a software can also be applied to develop imaging in echocardiography at this invention. Since left ventricle function is not just an elastic body movement but is a problem in fluid (blood)-wall interaction. Having used ultrasound method in echocardiography, fluid dynamic in the left ventricle is the major goal in FIG. **2**, as a fluid-wall interaction problem we are dealing to solve a system of partial differential equations one by continuum motions **27-34**, and the others are related to Navier-Stocks equations **35-40**. One of the major investigation is to solve discretely Navier-Stocks equations by experimental electrocardiogram data **41**, two main methods have been applied on electrocardiogram data at this invention **42-43**, that after a manipulating of these data, they can be considered as inputs of an algorithm where simulates blood through the left ventricle in LSDYNA software as a fluid dynamical simulation software **45**. Referring to FIG. **2** and FIG. **3** we give a detailed description of our method to solve of equations by using ultrasound method. One of the main attempts at this invention serves to study an arbitrary piece of the myocardium while is mechanically contracted, from the end-diastole to the end-systole. This finding can be posed to gain a good understanding of the local contractility, in contrast with chemical and electrical reasons of the cardiac muscle contraction. This assessment would be followed by a new notion of the power of myocytes resulting in the sufficient energy to the cardiac motion and deformation. This power was measured through 50 normal cases for each muscle volume samples in the myocardium at peak systolic. Case data included velocity, displacement, strain and strain rate. For instance, passing 0.2s from end diastole for a case in a piece with unit volume of his/her base lateral, 2.72w power is needed to get 4.5cm/s, 0.83cm, 25% and 1.85 1/s, in velocity , displacement, strain and strain rate respectively. This option is also considered as a remodeling index of the ventricles that as an example of this include due to vavular stenosis or regurgitation, coronary artery disease with associated decreased contractility and fibrosis leading to ventricular dilatation, genetic abnormalities resulting in hypertrophy, and progressive development of local fibrosis, severe hypertension, conduction abnormalities, etc. in fact this remodeling is a response to a problem with either the muscle itself or the environment in which it has to work and is an attempt to keep on fulfilling the heart's task-circulating the blood. However, since this is an abnormal situation with inherent mechanical disadvantages, in the long term, this will lead to irreversible damage to the muscle which evolves into ventricular dysfunction and heart failure. The early detection and follow-up of change in cardiac function and myocardial properties is thus of major importance.

The strain-stress relationship is a common concept used to investigate the mechanical properties of materials and has recently been applied to the analysis of cardiac tissue properties. The concept of strain ($\varepsilon$) corresponds to the deformation of an object as a function of an applied force stress (s). It represents the precentage of change of the unstressed dimension after the application of stress. And hold for both expansion (positive strains) and compression (negative strains). Considering a one dimensional object. The possible stress deformations are lengthening and shortening. In this case the strain can be expressed as the difference between the original length ($L_0$) and the length (L) after deformation. Normalized by the original length ($L_0$). As shown in the following equation:

$$\varepsilon = \frac{L - L_0}{L_0} \tag{1}$$

As shown, the strain is a dimensionless parametr and, by convention, positive correspond to lengthening and negative strains to shortening.

In some cases the length of the object is known during the deformation process, in that case we can define the instantaneous strain as:

$$\varepsilon(t) = \frac{L(t) - L(t_0)}{L(t_0)} \tag{2}$$

Where (L(t)) is the length at a given instant (t) and ($L(t_0)$) is the original length.

The strain rate is measurement of the rate of deformation and corresponds to the velocity of the deformation process. Taking into account equation (1) and (2) the latter definition, the instantaneous strain rate $\dot{\varepsilon}(t)$ is expressed as the temporal strain derivative:

$$\dot{\varepsilon}(t) = \frac{d\varepsilon(t)}{dt} = \frac{dL(t)}{dt.L(t)} = \frac{L'(t)}{L(t)} \tag{3}$$

With $L'(t)$ being the rate of deformation and L(t) the instantaneous length. The strain rate units are $s^{-1}$. Compared with the rate of deformation unites m. $s^{-1}$. As a conclusion, both strain rate and strain are closely related and they can be derived one from each other as we will discuss later.

Different approaches have been proposed to calculate the strain and strain rate parameters; some preliminary studies have been carried out to evaluate their precision and clinical usefulness. Two main methods have been proposed: the Crosscorrelation method and the velocity gradient method. The Cross-correlation method is based on the principles of elastography where two consecutive radiofrequency signals are compared to extract information about the tissue elasticity properties. Considering two consecutive radiofrequency signals applied to a tissue under deformation, the resultant backscatter signals received have similar patterns, except for a temporal shift related to the actual deformation. Cross-correlation analysis provides the temporal shift or delay introduced due to the object motion. From this analysis, different parameters such as the change in distance, local motion, velocities, etc., can be derived. The principal limitation of this technique is its high computational cost and the need of very high temporal resolution to avoid noisy estimates. That limits its application to M-mode acquisition. On the other hand, Fleming et al and Uematsu et al . introduced the concept of myocardial velocity gradient as an indicator of local contraction and relaxation. This concept is directly related to the already formulated strain rate parameter (3) under the assumption of linear and uniform strain (homogeneous, isotropic and incompressible material) as can be shown in the following expression:

$$\dot{\varepsilon}(t) = \frac{d\varepsilon(t)}{dt} = \frac{dL(t)}{dt \cdot L(t)} = \frac{L'(t)}{L(t)} \approx \frac{v_1(t)}{L(t)} - \frac{v_2(t)}{L(t)}.$$
(4)

where $v_1(t)$ and $v_2(t)$ are the local instantaneous velocities at two myocardial points separated an $L(t)$ distance. This formulation allows to compute myocardial strain rate as the spatial gradient of myocardial velocities, which can be obtained by Doppler echocardiography. Since the computational load is not high it can be implemented as a post-processing step after acquiring Doppler Tissue images or in real-time from digitally stored tissue velocity information . The axial natural strain component can be calculated from the strain rate curve time-integration expressed by:

$$\varepsilon(t) = \int_{t_0}^{t} \dot{\varepsilon}(t) dt \qquad (5)$$

When using myocardial motion and deformation to assess (dys-) function, it is important to understand the relation between intrinsic power and the resulting motion and deformation. In fact by roughly speaking, how much mechanical power is needed to make these dynamical deformations for each cardiac muscle volume? This matter would give the best information of cardiac function. In summary, the main factors influencing to the cardiac function mechanically are: 1) Regional myocardial motion and deformation. 2) Intrinsic power, i.e. the sufficient energy force developed by myocardium, resulting in motion and deformation. Having applied ViVid7 setting, for 50 normal cases, we got velocity, displacement, strain and strain rate at peak systolic, for each part of the myocardium, i.e. at the base, Middle and Apical of Septal, Lateral, Anterior and inferior FIG. **10**.

And using these data and applying them in the "Power formula", we could compute intrinsic power values at peak systolic, for each local part of the myocardium where these values have been presented at the diagrams in FIG. **11**, for instance, the value of this new option at the base septal for a case is 6.5w that means, 6.5w is needed to make a necessary motion and deformation of this part at the peak systolic. And with removing the general motion of the heart, this value for that case at the base septal is less than 6.5w where this states, tethering helps that piece of myocardium to handle the matter.

We pick a cardiac muscle volume sample, for instance from the basal septum and follow-up its motion and deformation during the end diastole to the end systole FIG **12**.

Passing time" t" after the end diastole, radial strain rate for the fiber AB is:

$$strain\ rate = V_B(t) - V_A(t)/L_{AB}(t)$$

Let $D_t$ and $W_t$ are displacement and velocity of the above muscle volume element at time t.

By classical mechanic we set :

$$D_t + L_{AB}(t) - L(t_0) = \frac{1}{2}a_t t^2 + W_t t$$

$t_0$ is the time and $L(t_0)$ is the length of the fiber AB respectively at the end diastole.

Having used the 1-D deformation definition we rewrite the above formula by the following way:

$$\frac{D_t}{t} + \frac{L_{AB}(t)}{t} - \frac{L(t_0)}{t} = \frac{1}{2}a_t t + W_t$$

By radial strain rate and strain of the fiber AB at time "t" we have:

$$\frac{D_t}{t} + strain\ rate(t).(strain(t) + 1)L(t_0) = \frac{1}{2}a_t t + W_t$$

Thus we can reformulate $a_t$ in below:

$$a_t = 2\frac{D_t}{t^2} + 2 strain\ rate(t).\frac{(strain(t)+1)L(t_0)}{t}$$
$$-\frac{2W_t}{t} \qquad (6)$$

Now if μ be the density and Volume(t) (8) be the volume respectively of our muscle volume sample after the contraction during the time "t". the power which is needed to result in this motion and deformation at time "t" would be FIG. **13**:

$$P(t) = \mu.Volume(t).a_t.D_t/t$$

Decreased or increased in power, for a muscle volume sample can be caused a maker of myocardial ischemia, LV dysfunction, or LV hypertrophy FIG. **4, 59**. Inasmuch as intrinsic power has simultaneously been contributed by some codependent factors such as motion and deformation FIG. **2, 18,19**, and **22**, it is realized as a remodeling index to keep up normal conditions. For instance failure to achieve adequate wall thickening (at radial deformation) and lengthening/shortening (at longitudinal deformation) of a fiber can be made by power. Having increased in wall thickening of a piece of left ventricle, motivate us to inhibit widely dilate with a hypertrophic wall, by a reduction in power of that part. Particularly when power tends to zero for a muscle volume, it may be a factor that shows myofibers of that section, have died. So it would be a criterion, where we can make use of stern cell method to improve cardiac? FIG. 1 and 5

On the other hand, this power shows its influence on red blood cell flow by change in volume FIG. **3, 27**, In fact when a unit volume of red blood cell is dealing with a piece of ventricles, has a simple description of its change in volume, i.e. it can calculate by:

$\Delta V(t)$= Local change in volume of RBC with unit volume from end diastole to end systole , where

**$\Delta V(t)$= (Radial deformation(t))x(Longitudinal deformation(t)) x Displacement(t)** (8)

From this point of view, one can locally compute myocardial pressure which is a very valuable measure for many normal and pathological conditions of the heart, particularly right ventricular function and interrelationship of the right and left heart. This pressure is directly measured by the following way:

**Local myocardial pressure (t) =$P(t).t/\Delta V(t)$.**
**(9)**

If we follow myofiber's transactions from the end diastole to the end systole, they would show us curves FIG. **2, 23**, that intrinsic power is made to move myofibers on these bands.
Referring to FIG. **4**, the present invention includes algorithms that are applied to softwares **50** and **53** to follow a muscle volume sample in the myocardium and its interaction with a red blood cell within the LV contractions. Experimental samples are evaluated by electrocardiogram data and mathematical formulations will be also applied on these data to disclose images of a fiber modeling of the left ventricle motions and blood through it. Referring to a ventricular disease in terms of its electrocardiogram data, we can also do the same work to give a fiber modeling for this disease **59**, that it would basically be hard to diagnosis it by echo techniques. By a technical manipulating of electrocardiogram data therein as inputs of our algorithms we can do cardiac surgeries virtually **53** and **60**. Having made the beneficiary used of fiber modeling of the LV where had been modeled by echocardiography tool and referring to FIG. **4**, and **51, 52** we can make a program to virtual reality in cardiac surgery and follow new techniques and repairs for cardiac surgical tasks.

Due to FIG. **5**, the best mathematical modeling at this invention was the main goal where it can be applied on an elastic polymer for an artificial left ventricle **64** and **65**. In fact dynamical properties of this polymer are made by representing elastic tubes (as myofibers in the left ventricle) where have been equipped to micro actuators in their nodal points to charge enough powers **66-69**.

The present study has been described in terms of a mathematical modeling that all mathematical formulations are stated from ultrasound methods in compared with the other cardiac imaging tools.

Disclosure:

There is no disclosure.